\documentclass[doublecol]{epl2} 
\pdfoutput=1

\usepackage{color}
\usepackage{amsmath}
\usepackage{amssymb}

\usepackage{xspace}

\title{Observation of an isotropic superconducting gap at the Brillouin zone center of Tl$_{0.63}$K$_{0.37}$Fe$_{1.78}$Se$_2$}
\shorttitle{Isotropic superconducting gap at the Brillouin zone center of Tl$_{0.63}$K$_{0.37}$Fe$_{1.78}$Se$_2$} 

\author{X.-P. Wang\inst{1,2} \and P. Richard\inst{1} \and X. Shi\inst{1} \and  A. van Roekeghem\inst{1,3} \and Y.-B. Huang\inst{1} \and  E. Razzoli\inst{2} \and T. Qian\inst{1} \and E. Rienks\inst{4} \and S. Thirupathaiah\inst{4} \and H.-D. Wang\inst{5} \and C.-H. Dong\inst{5} \and M.-H. Fang\inst{5} \and M. Shi\inst{2} \and H. Ding\inst{1}\thanks{\email{dingh@iphy.ac.cn}}}
\shortauthor{X.-P. Wang \etal}

\institute{                    
  \inst{1} Beijing National Laboratory for Condensed Matter Physics, and Institute of Physics, Chinese Academy of Sciences, Beijing 100190, China\\
  \inst{2} Paul Scherrer Institut, Swiss Light Source, CH-5232 Villigen PSI, Switzerland\\
  \inst{3} Centre de Physique Th\'{e}Žorique, Ecole Polytechnique, CNRS-UMR7644, 91128 Palaiseau, France\\
  \inst{4} Helmholtz-Zentrum Berlin, BESSY, D-12489 Berlin, Germany\\
  \inst{5} Department of Physics, Zhejiang University, Hangzhou 310027, China
}

\abstract{
We performed a high-resolution angle-resolved photoemission spectroscopy study on superconducting (SC) Tl$_{0.63}$K$_{0.37}$Fe$_{1.78}$Se$_2$ ($T_c=29$ K) in the whole Brillouin zone (BZ). In addition to a nearly isotropic $\sim$ 8.2 meV 2-dimensional (2D) SC gap ($2\Delta/k_BT_c\sim7$) on quasi-2D electron Fermi surfaces (FSs) located around M$(\pi,0,0)$-A$(\pi,0,\pi)$, we observe a $\sim 6.2$ meV isotropic SC gap ($2\Delta/k_BT_c\sim5$) on the Z-centered electron FS that rules out any $d$-wave pairing symmetry and rather favors an $s$-wave symmetry. All isotropic SC gap amplitudes can be fit by a single gap function derived from a local strong coupling  approach suggesting an enhancement of the next-next neighbor exchange interaction in the ferrochalcogenide superconductors. 
}

\begin{document}

\maketitle

One of the most important and intensively debated issues in Fe-based superconductivity is whether the superconducting (SC) pairing is caused by itinerant magnetic fluctuations enhanced by Fermi surface (FS) quasi-nesting or by local magnetic superexchange interactions \cite{Richard_PoPP2011}. For most Fe-based superconductors, which have quasi-nested hole and electron FSs separated by the antiferromagnetic (AF) vector, both the weak-coupling itinerant  and the strong-coupling local approaches predict a similar sign-reversed $s$-wave pairing symmetry. However, for the newly discovered ferrochalcogenide superconductor A$_x$Fe$_2$Se$_2$ (A = K, Rb, Cs, Tl) \cite{GuoJG_PRB2010,FangMH_EPL2011},  which does not have hole FS pocket at the Brillouin zone (BZ) center as observed by angle-resolved photoemission spectroscopy (ARPES) \cite{Wang_EPL2011,FengDL_NM2011,ZhouXJ_PRL2011,QianT_PRL2011}, most calculations \cite{WangFa_EPL2011,Mazin_PRB2011,Scalapino_PRB2011,Kontani_PRB2011,Das_PRB2011,ZhouTao_arxiv2012} based on the weak-coupling approach predict a sign reversal in the gap between the electron FS sheets located at M$(\pi,0)$ and $\widetilde{\textrm{M}}$$(0,\pi)$ (here defined in the 1 Fe/unit cell notation) naturally described by a $d$-wave symmetry. On the other hand, the strong-coupling approach still favors a $s$-wave symmetry \cite{HuJP_PRL2008,HuJP_PRX2011,HuJP_SR2012,YuR_arxiv2011,HuJP_SR2012,ZhouYi_EPL2011, HuJP_PRX2012}. 

Although the observation by ARPES of an isotropic SC gap on the electron FSs at the zone boundary \cite{Wang_EPL2011,FengDL_NM2011,ZhouXJ_PRL2011} supports the $s$-wave scenario, it cannot completely rule out a $d$-wave symmetry which in principle can be nodeless if the nodal lines do not intersect any FS. Luckily, there is a 3-dimensional (3D) electron FS pocket centered around Z$(0,0,\pi)$ as previously reported by ARPES \cite{FengDL_NM2011,LiuZH_PRL109}, which can be used to distinguish $d$-wave from  $s$-wave symmetry since the $d$-wave symmetry would create nodes along a pair of perpendicular directions on this FS pocket \cite{Kontani_PRB2011,Das_PRB2011,ZhouTao_arxiv2012}. Therefore, the momentum ($k$)-dependence of the SC gap on this electron FS is crucial in determining the pairing symmetry and mechanism of this superconductor and possibly the entire family of Fe-based superconductors. 

In this letter, we report high-resolution ARPES measurements on SC Tl$_{0.63}$K$_{0.37}$Fe$_{1.78}$Se$_2$ ($T_c=29$ K) in the whole 3D BZ by tuning the incident photon energy. We observe a nearly isotropic and nearly $k_z$-invariant $\sim$ 8.2 meV SC gap on electron FS pockets located around M, in agreement with our previous report on the same material using a He discharge lamp \cite{Wang_EPL2011}. More significantly, a nearly isotropic SC gap with a smaller size of $\sim$ 6.2 meV is observed on the 3D electron FS pocket around Z, without obvious variation along $k_z$. This smaller gap closes at the same temperature as the larger gap near M \cite{Wang_EPL2011}, suggesting a same pairing channel for the two different FSs. Our observation of a nodeless SC gap structure for the whole \emph{k}-space rules out the \emph{d}-wave pairing and rather supports a \emph{s}-wave symmetry. Interestingly, the global gap structure in this material can be described by a single gap function derived from a strong-coupling approach \cite{HuJP_PRX2011,HuJP_SR2012}.

High-quality single crystals of  Tl$_{0.63}$K$_{0.37}$Fe$_{1.78}$Se$_2$ ($T_c^{onset}=29.1$ K; $T_c^{mid}=28.6$ K; $T_c^{zero}=27.5$ K) were grown by the Bridgeman method \cite{FangMH_EPL2011}. The precise composition was determined using energy dispersive X-ray spectrometry (EDXS). The lattice parameters $a=3.85$ \AA\xspace and $c=14.05$ \AA\xspace were determined by fitting XRD data. ARPES measurements were performed at the 1-cubed ARPES end-station of BESSY and at Swiss Light Source beamline SIS using a VG-Scienta R4000 electron analyzer with photon energy ranging from 18 to 63 eV. The angular resolution was set to 0.2$^{\circ}$ and the energy resolution to 5-7 meV for SC gap measurements at $T=0.9$ K. The band structure and temperature-dependent SC gap sizes were recorded with 10-15 meV energy resolution. All samples were cleaved \emph{in situ} and measured in a working vacuum better than 5$\times$10$^{-11}$ torr. For convenience, we describe all the results using the 1 Fe site/unit cell (or unfolded) notation and use $c'=c/2=7.025$ \AA. In the following, we also use $\bar\Gamma$ and $\bar{\textrm{M}}$ to refer to any point along the $\Gamma(0,0,0)$-Z$(0,0,\pi)$ axis and M$(0,\pi,0)$-A$(0,\pi,\pi)$ axis, respectively.

\begin{figure}[!t] \includegraphics[width=8.5cm]{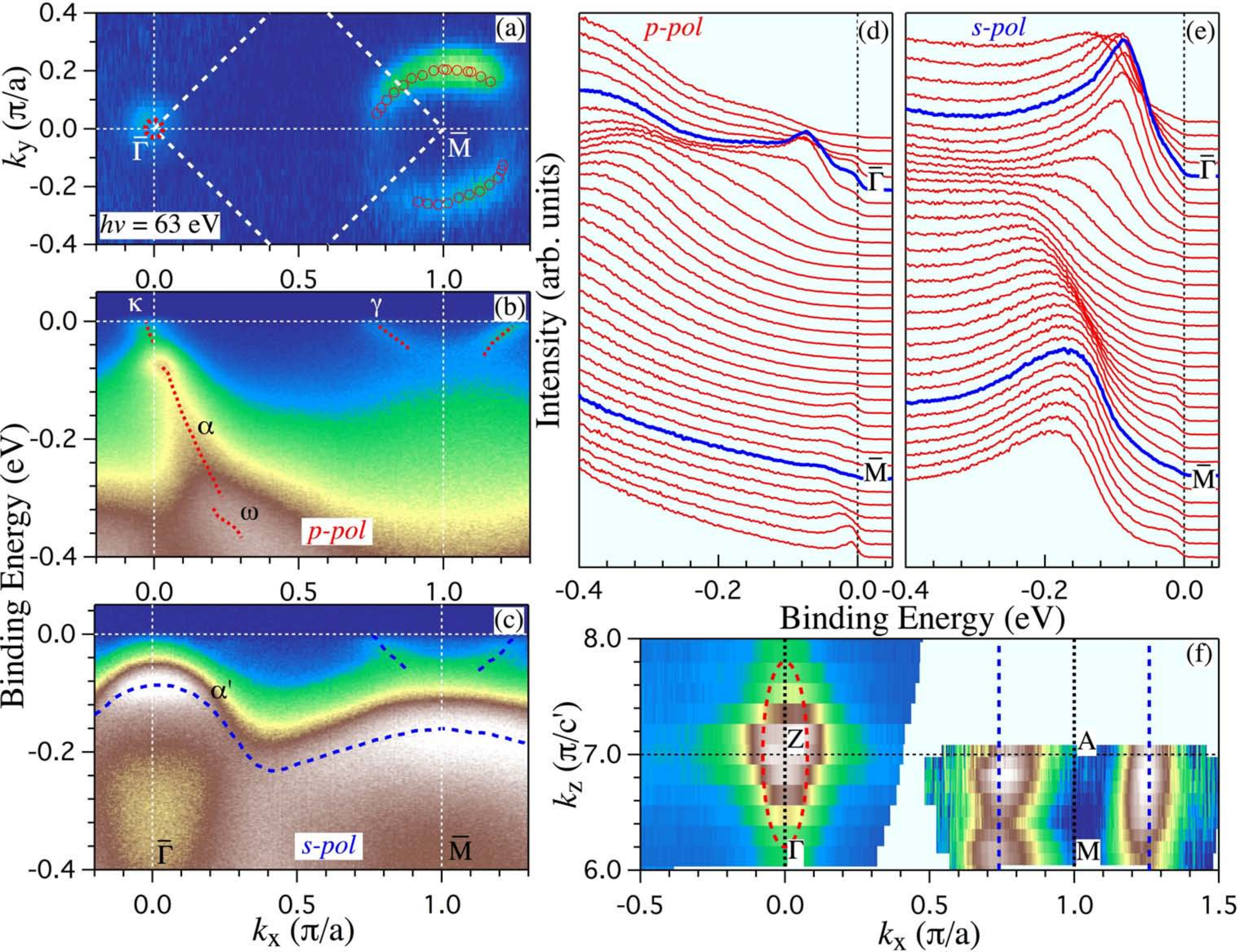} 
\caption{\label{FS_plot}(Color online) (a) ARPES FS intensity map of Tl$_{0.63}$K$_{0.37}$Fe$_{1.78}$Se$_2$ ($\pm 5$ meV integrated window) recorded in the normal state (35 K) with 63 eV photons. Open circles and filled triangles correspond to $k_F$ locations of the $\gamma$ and $\kappa$ bands, respectively. (b) ARPES intensity plot ($h\nu=63$ eV) for a cut along the $\bar\Gamma$-$\bar{\textrm{M}}$ direction recorded at 35 K with a $p$ polarization. Guides to the eye are plotted for the various bands observed. (c) Same as (b) but using $s$-polarized photons. (d)-(e) EDCs corresponding to the cuts in (b) and (c), respectively. (f) ARPES intensity plot in the $k_z$-$k_x$ plane. The red and blue dashed lines indicate the $k_F$ locations.} 
\end{figure}

In Fig. \ref{FS_plot}(a), we show an ARPES FS intensity mapping of Tl$_{0.63}$K$_{0.37}$Fe$_{1.78}$Se$_2$ recorded in the normal state (35 K) with 63 eV photons. As with previous results \cite{QianT_PRL2011,Wang_EPL2011,ZhouXJ_PRL2011,FengDL_NM2011,LiuZH_PRL109}, the dominant feature is an almost circular electron FS centered at the $\bar{\textrm{M}}$ point. An additional weak intensity spot is also detected at $\bar\Gamma$. We display in Figs. \ref{FS_plot}(b) and (c) the ARPES cuts recorded along $\bar\Gamma$-$\bar{\textrm{M}}$ with $p$-polarized and $s$-polarized light, respectively. As with the corresponding sets of energy distribution curves (EDCs) given in Figs. \ref{FS_plot}(d) and (e), they indicate that this feature has an electron-like dispersion that will hereafter be called the $\kappa$ band. In agreement with a previous study on (Tl,Rb)$_{y}$Fe$_{2-x}$Se$_2$ \cite{LiuZH_PRL109}, our light polarization analysis indicates that the $\kappa$ band is enhanced with $p$ polarization as compared with $s$ polarization. 

The out-of-plane momentum $k_z$ can be approximated in ARPES experiments by tuning the photon energy over a wide range. The estimated values of $k_z$ are given by $k_z=\sqrt{2m_e/\hbar^2[(h\nu-\phi-E_B)\cos^2\theta+V_0]}$, where $\phi$ is the work function, $V_0$ is the inner potential of the sample (estimated to 12 eV for this material from the periodic dispersion along $k_z$), and $\theta$ is the emission angle \cite{Hufner_1995}. In Fig. \ref{FS_plot}(f), we show the intensity mapping in the $k_x-k_z$ plane of Tl$_{0.63}$K$_{0.37}$Fe$_{1.78}$Se$_2$ over one half-cycle. According to our analysis, $h\nu$ = 18 and 30 eV are close to the $\Gamma$ and Z points, respectively. While the FS centered at $\bar{\textrm{M}}$ does not disperse with $k_z$, strong $k_z$ modulations are observed for the small 3D FS pocket centered at the Z point, as pointed out previously \cite{FengDL_NM2011,LiuZH_PRL109}.

\begin{figure}[!t] \includegraphics[width=8.5cm]{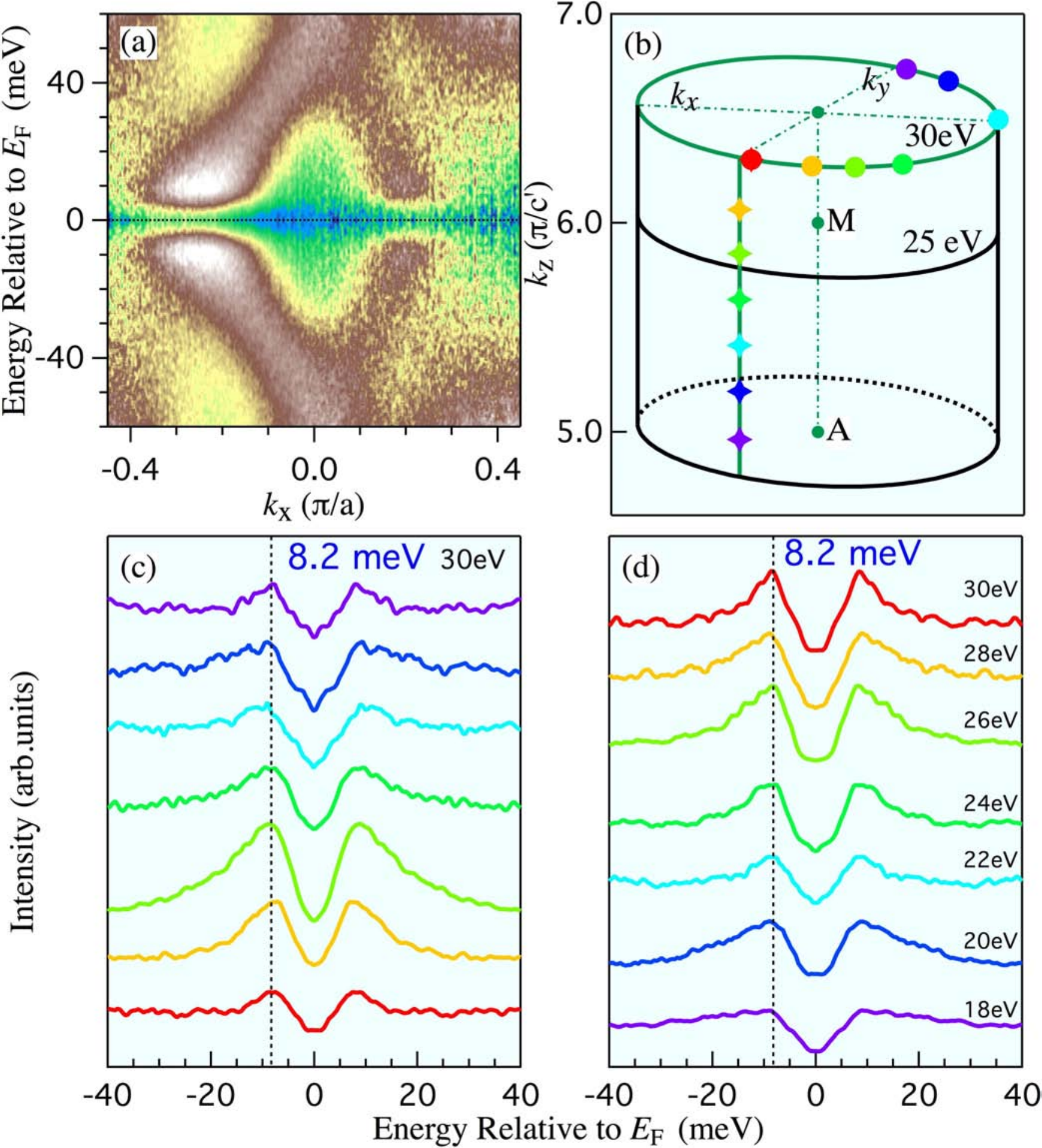} 
\caption{\label{gamma_band}(Color online) (a) Symmetrized ARPES intensity plot of the $\gamma$ band for $h\nu=30$ eV. (b) Schematic 3D $\gamma$ FS and $k_F$ locations of the EDCs used for the SC gap measurements. (c) Symmetrized ARPES spectra in the SC state (0.9 K) measured with $h\nu=30$ eV at various $k_F$ points indicated by circles in (b). (d) Symmetrized EDCs at various $k_F$ positions along the $k_z$ axis indicated in (b) with diamonds. The color of the EDCs in (c)-(d) refers to the symbol of the $k_F$ indicators in (b). Dashed lines are guides to the eye.} 
\end{figure}

To check whether the SC gap follows the same modulations as the Fermi momentum $k_F$, we performed high-resolution SC gap measurements at different $k_z$ values. The results for the $\gamma$ band are shown in Fig. \ref{gamma_band}. We display in Fig. 2(a) the intensity plot of a cut recorded at $0.9$ K with $h\nu=30$ eV that crosses the $\bar{\textrm{M}}$, which has been symmetrized with respect to $E_F$ following a common practice, to approximately remove the contribution of the Fermi-Dirac function at the $k_F$ points in the SC state. A SC gap opening with a strong coherent peak and the bending back in the dispersion characterizing the SC state are clearly observed. Using this procedure, one can identify precisely the minimum gap location, which corresponds to $k_F$. The location of the $k_F$ points on the $\gamma$ FS where the SC gap has been measured using 30 eV photons is given by circles in Fig. \ref{gamma_band}(b). The corresponding symmetrized EDCs, shown in Fig. \ref{gamma_band}(c), indicate a nearly isotropic SC gap $\Delta$, defined by half the distance between the two peaks, which averages to 8.2 meV. This value is consistent with our previous study on the same material using the I$\alpha$ line of a Helium discharge lamp \cite{Wang_EPL2011}. In Fig. \ref{gamma_band}(d), we show additional symmetrized EDCs recorded with photon energies ranging from 18 eV to 32 eV, which covers the $k_z$ range indicated in Fig. \ref{gamma_band}(b). Our results indicate that the gap size associated with the $\gamma$ band is quite $k_z$-insensitive. 

\begin{figure}[!t] \includegraphics[width=8.5cm]{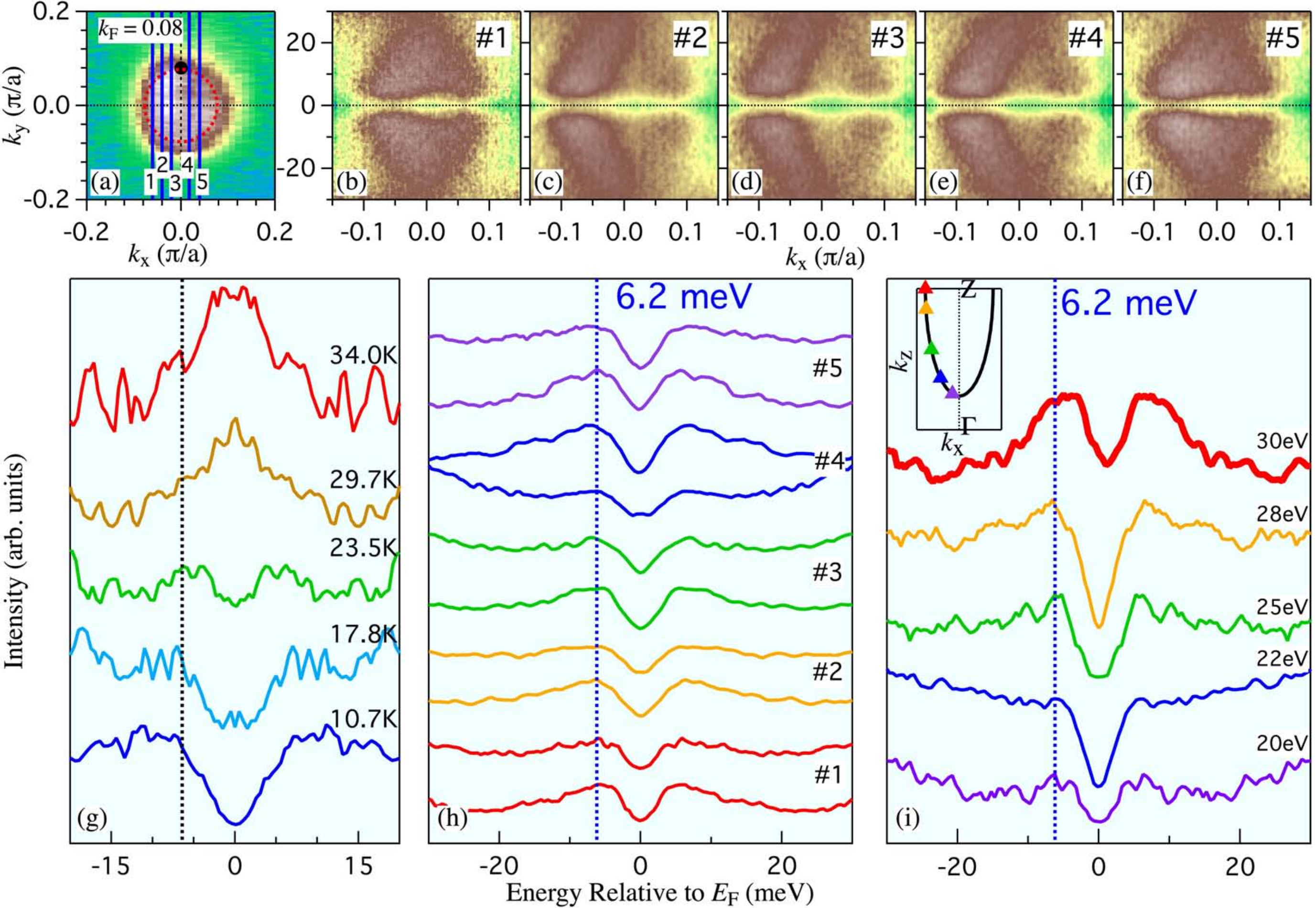}
\caption{\label{kappa_band}(Color online) (a) ARPES intensity map around Z$(0,0,\pi)$ with photon energy 30 eV. The  red dashed line indicates the FS position with $k_F\sim0.08\pi/a$. (b)-(f) Symmetrized ARPES intensity plot recorded in the SC state (0.9 K) along cuts \#1 to \#5 indicated by blue lines in (a). (g) Temperature dependence of symmetrized EDC spectra at $k_F$. The black dot in (a) shows the $k_F$ location of the EDCs. (h)-(i) Symmetrized EDCs in the SC state (0.9 K) measured at various $k_F$ points along the $\kappa$ FS in the $k_z$=$\pi$ plane, and along $k_z$, respectively. The $k_F$ positions in (i) are indicated in the inset. Dashed lines are guides to the eye. } 
\end{figure}

We now turn our attention to the 3D $\kappa$ FS pocket, for which we performed similar measurements and analysis. The results are reported in Fig. \ref{kappa_band}. In panel (a) we show the FS intensity map obtained at 30 eV around the Z point ($k_z\sim\pi$). A series of symmetrized intensity plots along the cuts identified in Fig. \ref{kappa_band}(a),  is given in panels (b)-(f). These cuts illustrate that a SC gap opens everywhere along the $k_z=\pi$ section of the $\kappa$ FS. The temperature evolution of the $k_F$ EDCs given in panel (g) shows that the gap fills up above the SC transition ($T_c=29$ K), confirming its SC nature. We note that this behavior is reversible upon thermal cycle. The symmetrized EDCs corresponding to the $k_F$ positions of the cuts \#1 to \#5 in Fig. \ref{kappa_band}(b)-(f) are displayed in Fig. \ref{kappa_band}(h). Although the precise determination of the SC gap size would allow larger anisotropy ($\leq$ 20\%) than for the $\gamma$ band, the SC gap of the $\kappa$ band is also nodeless and quite isotropic in this particular $k_z$ plane. The averaged SC gap value for $h\nu=30$ eV is approximately 6.2 meV, which is smaller than that of the $\gamma$ FS but still in the strong coupling regime ($2\Delta/k_BT_c\sim5$). To investigate possible SC gap modulation along the dispersive $k_z$ direction, we tuned the photon energy between 20 eV and 30 eV. The corresponding symmetrized EDCs are displayed in Fig. \ref{kappa_band}(i). Surprisingly, the SC gap size is robust against the strong modulations of the FS, indicating a full 3D SC gap with $s$-wave symmetry.

\begin{figure}[!t] \includegraphics[width=8.5cm]{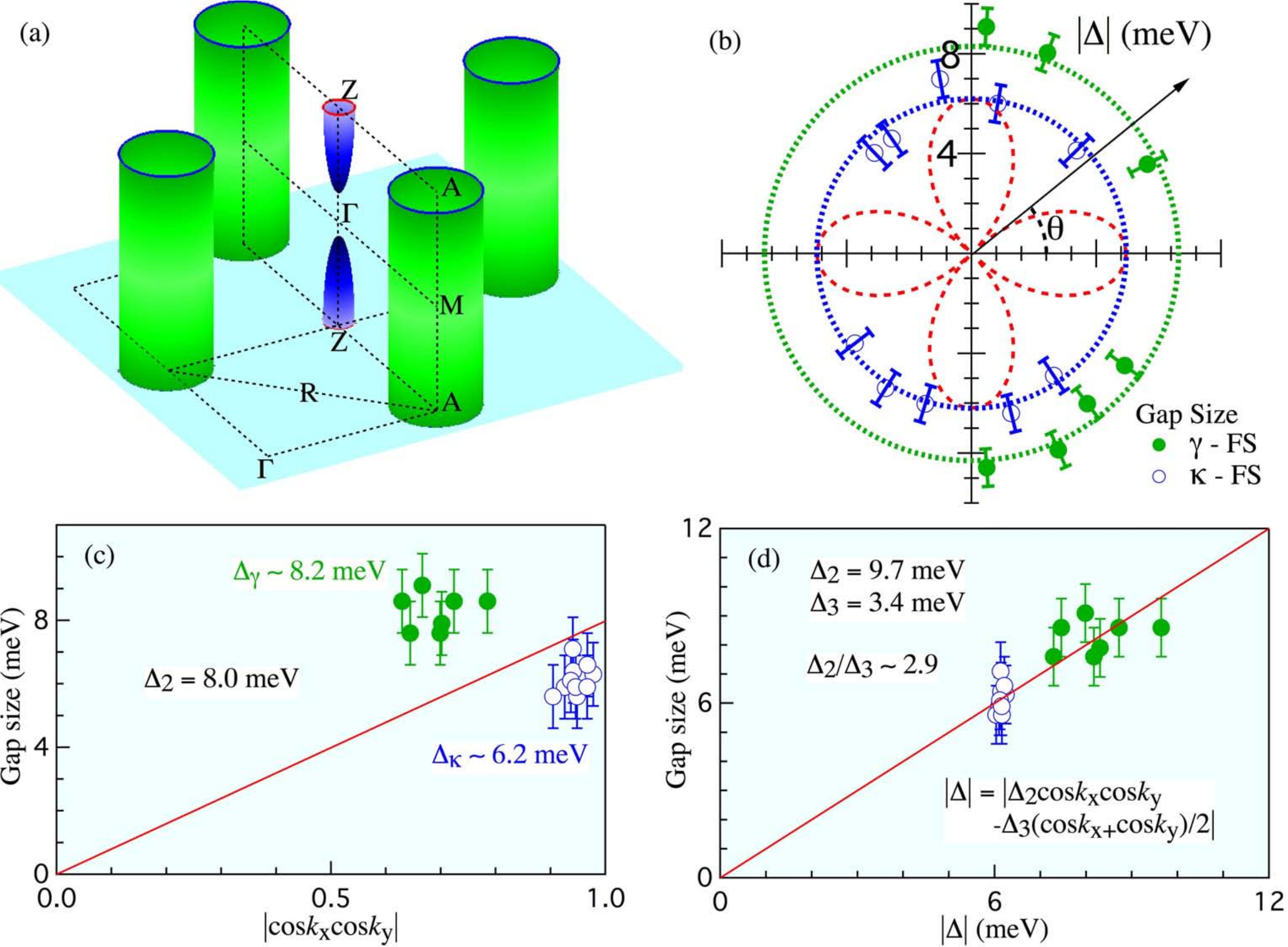} 
\caption{\label{SC_gap}(Color online) (a) 3D representation of the FS of Tl$_{0.63}$K$_{0.37}$Fe$_{1.78}$Se$_2$. (b) Polar representation of the SC gap associated with the $\kappa$ and $\gamma$ FSs at 0.9 K and $h\nu=30$ eV (near $k_z\sim\pi$). The red dashed line indicate a $d$-wave profile. (c)-(d) SC gap values at 30 eV plotted as a function of $|\cos k_x\cos k_y|$ and $|\Delta_2\cos k_x\cos k_y-\Delta_3(\cos 2k_x+\cos 2k_y)/2|$, respectively. } 
\end{figure}

Our main results on the SC gap in Tl$_{0.63}$K$_{0.37}$Fe$_{1.78}$Se$_2$ are summarized in Fig. \ref{SC_gap}. The FS topology illustrated in Fig. \ref{SC_gap}(a), with the absence of hole FS \cite{QianT_PRL2011,Wang_EPL2011,FengDL_NM2011,ZhouXJ_PRL2011,LiuZH_PRL109}, seems inconsistent with electron-hole quasi-nesting pairing mechanisms. The complete SC gap structure, which is directly shaped by the SC pairing mechanism, can narrow further the list of the pairing candidates. In Fig. \ref{SC_gap}(b) we show the polar representation of the SC gap measured for the $\kappa$ and $\gamma$ FSs. As mentioned above, these gaps are nodeless and isotropic within experimental uncertainties, and they show no obvious variation along $k_z$. While the isotropy of the SC gap around $\bar{\textrm{M}}$ does not directly exclude a global $d_{x^2-y^2}$-wave symmetry with nodes along the $k_x=\pm k_y$ directions, the one of the $\kappa$ FS pocket elongated along the $\Gamma$-Z direction is a strong evidence against such pairing symmetry. On the contrary, it is rather consistent with bulk-sensitive nuclear magnetic resonance \cite{YuWQ_2011} and specific heat measurements \cite{ZengB_PRB2011} which suggest the absence of node in the SC gap structure. One could argue that an $s$-wave pairing on the 3D FS pocket at Z may coexist with a $d$-wave pairing on the large 2D FS around $\bar{\textrm{M}}$ and that the different pairings are only weakly coupled. However, such an ``hybrid" pairing scenario usually has different pairing strengths and different transition temperatures above which the SC gap closes. Our observation of the same temperature dependence on both FSs does not support this scenario.

An $s$-wave symmetry can be derived from various theoretical approaches. The conventional $s_{\pm}$ symmetry with a sign reversal between $\Gamma$ and M can be obtained from an effective $J_1$-$J_2$ model \cite{HuJP_PRL2008,HuJP_PRX2011,HuJP_SR2012,LuX_PRB2012, HuJP_PRX2012,YuR_arxiv2011} while the valence band Suhl-Kondo effect can lead to an $s_{++}$ pairing symmetry, without any phase sign change \cite{Kontani_PRB2011}. Weak coupling scenarios including orbital fluctuations \cite{Kontani_PRB2011} and charge fluctuations \cite{ZhouS_PRB2011} may also lead to an $s_{++}$ paring symmetry, while a ``bonding-antibonding'' $s_{\pm}$ symmetry with a sign reversal between two concentric M-centered FS pockets without any node in the whole $k$-space was proposed based on a spin fluctuations approach \cite{Mazin_PRB2011}. Even though our $k$-sensitive but phase-insensitive ARPES measurements do not allow us to identify directly which of the $s$-wave pairing symmetries is the correct one, the relative gap size on the various FSs may give some hints on the relevant pairing mechanism. 

Recently, a careful ARPES study of the SC gap in the cousin FeTe$_{0.55}$Se$_{0.45}$ ferrochalcogenide superconductor reported nearly isotropic SC gaps with a larger gap at the M point \cite{MiaoH_PRB2011}. While the SC gap of most ferropnictide superconductors is roughly consistent with a $|\cos k_x\cos k_y|$ gap function \cite{Nakayama_EPL2009,Nakayama_PRB2011,LiuZH_PRB2011} that can be derived from a $J_1$-$J_2$ model \cite{HuJP_PRL2008}, where $J_1$ and $J_2$ characterize the nearest- and next-nearest-neighbor magnetic exchange interaction couplings, the global SC gap structure in FeTe$_{0.55}$Se$_{0.45}$ rather follows a $|\Delta_2\cos k_x\cos k_y-\Delta_3(\cos 2k_x+\cos 2k_y)/2|$ function consistent with a non-negligeable $J_3$ constant \cite{HuJP_PRX2011}. In order to check if either of these gap functions applies to Tl$_{0.63}$K$_{0.37}$Fe$_{1.78}$Se$_2$, we show in Figs. \ref{SC_gap}(c) and \ref{SC_gap}(d) fits of our SC gap data to the $J_1-J_2$ and $J_1-J_2-J_3$ gap functions, respectively. While the data are clearly inconsistent with the former model, the latter works reasonably well. From the fit to the $|\Delta_2\cos k_x\cos k_y-\Delta_3(\cos 2k_x+\cos 2k_y)/2|$ gap function, we obtain the global SC gap parameters $\Delta_2=9.7$ meV and $\Delta_3=3.4$ meV. Their ratio gives 2.9, which is not much different but smaller than the $\sim 3.7$ ratio obtained in FeTe$_{0.55}$Se$_{0.45}$ \cite{MiaoH_PRB2011}, suggesting a stronger $J_3$ coupling constant. Within the $J_1-J_2-J_3$ model, an increase in the $J_3$ strength (or a decrease in the $J_2/J_3$ ratio) is consistent with the observation of a strong Se 4$p_z$ orbital component for the $\kappa$ band in (Tl,Rb)$_{y}$Fe$_{2-x}$Se$_2$ \cite{LiuZH_PRL109}, which favors the Se-bridged superexchange interaction that can enhance $J_3$. However, we notice that the latter fit has a good degree of uncertainty, and more precise measurements are needed in order to make a definite claim.

In summary, we fully characterized the SC gap of Tl$_{0.63}$K$_{0.37}$Fe$_{1.78}$Se$_2$ over the whole BZ, including the small 3D electron pocket centered at Z. All gaps are nodeless and do not show significant momentum dependence, thus ruling out any $d$-wave pairing scenario in this system. Moreover, all SC gap data can be fit with a single gap function derived from a strong-coupling approach.

\acknowledgments

We are grateful to J.-P. Hu for useful discussions. This work is supported by the Chinese Academy of Sciences (grant No. 2010Y1JB6), the Ministry of Science and Technology of China (grants No. 2010CB923000, No. 2011CBA0010), and the Nature Science Foundation of China (grants No. 10974175, No. 11004232, and No. 11050110422). This work was supported by the Sino-Swiss Science and Technology Cooperation (project no. IZLCZ2 138954). This work was partly performed at the Swiss Light Source, Paul Scherrer Institut, Villigen, Switzerland, and at BESSY, Helmholtz Zentrum, Berlin, Germany.

\bibliographystyle{eplbib}
\bibliography{REF245}

\end{document}